\documentclass{article}

\usepackage{PRIMEarxiv}

\usepackage[utf8]{inputenc} 
\usepackage[T1]{fontenc}    
\usepackage{hyperref}       
\usepackage{url}            
\usepackage[numbers]{natbib} 
\usepackage{booktabs}       
\usepackage{amsmath}        
\usepackage{amsthm}         
\usepackage{float}          
\usepackage{amsfonts}       

\newtheorem{definition}{Definition}
\usepackage{nicefrac}       
\usepackage{microtype}      
\usepackage{lipsum}
\usepackage{fancyhdr}       
\usepackage{graphicx}       
\usepackage{fontawesome5}   
\usepackage{subcaption}     
\usepackage{listings}       
\graphicspath{{media/}}     

\title{Graphify: Automated Synthesis of Type-Safe Graph Backends via $O(S)$ GraphQL-to-Gremlin Transpilation}

\author{
    Johannes Graf \\
    Northwestern University \\
    Evanston, IL \\
    \texttt{johannesgraf2026@u.northwestern.edu}
    \thanks{Work conducted while at Cooperative State University Heidenheim. \\ 
    Source code: \url{https://github.com/Graf-J/Graphify}}
}

\begin{document}
\maketitle

\begin{abstract}
Graph databases offer unparalleled flexibility for managing interconnected data, yet the lack of strict schema enforcement often leads to runtime uncertainties and complex query development. 
This paper introduces Graphify, an end-to-end framework that enables developers to visually model graph data schemas and automatically synthesize a fully functional, type-safe backend. 
This paper proposes a formal mapping of GraphQL artifacts to the Gremlin traversal machine, supporting complete CRUD operations and arbitrarily nested queries. 
The system generates a transpiler capable of converting complex GraphQL requests into a single, optimized Gremlin query, including advanced features such as nested logical predicates, multi-key sorting, and pagination. 
At the core of the framework is a recursive state machine algorithm that processes GraphQL Abstract Syntax Trees (ASTs) with linear time complexity $O(S)$ relative to the number of selected fields. 
This paper demonstrates the practical efficiency and theoretical robustness of the approach through formal complexity analysis and empirical evaluation using the MovieLens 100k dataset. 
The result is a system that enables the generation of graph interfaces in minutes, bridging the gap between flexible graph storage and type-safe API consumption.
\end{abstract}


\section{Introduction}
In recent years, there has been a resurgence in the relevance of graph databases, owing to their inherent flexibility and high performance in managing interconnected data. 
This resurgence is driven by the growing demand for applications such as accurate recommendations for social networks, storing information networks such as citations among academic papers, or managing technological networks like electric power grids. 
Graph databases offer superior speed in executing complex queries to access interconnected data compared to classical relational databases \cite{Do2022}.
However, the possibilities of a standardized data schema and constraint definitions often are limited \cite{Jaroslav2015}. 
Therefore, the flexible data schema of graph databases can lead to uncertainties when querying for data, stemming from a lack of knowledge about the underlying data structure \cite{Canovas2013}. 

This paper explores property graph databases, which diverge notably from relational databases in their query language requirements, reflecting distinct functional needs. 
Unlike relational databases utilizing SQL, property graph databases lack a standardized method for information retrieval. 
Notable query languages in this context include Cypher \cite{Neo4jCypher}, Gremlin \cite{TinkerPopGremlin, TinkerPopGithub}, and GraphQL \cite{GraphQL, GraphQLSpec}. 
The following examples in Figure \ref{fig:query_examples} illustrate querying for a person named John using a graph database, showcasing the syntax disparities among the different query languages.

\begin{figure}[H]
    \centering
    \begin{minipage}[b]{0.35\textwidth}
        \begin{lstlisting}[language=SQL, basicstyle=\small\ttfamily]
MATCH (p:Person {name: 'John'})
RETURN p;
        \end{lstlisting}
        \centering\small(a) Cypher
    \end{minipage}
    \hfill
    \begin{minipage}[b]{0.25\textwidth}
        \begin{lstlisting}[language=Java, basicstyle=\small\ttfamily]
g.V()
 .hasLabel('Person')
 .has('name', 'John')
 .toList();
        \end{lstlisting}
        \centering\small(b) Gremlin
    \end{minipage}
    \hfill
    \begin{minipage}[b]{0.3\textwidth}
        \begin{lstlisting}[language=HTML, basicstyle=\small\ttfamily]
query {
  persons(name: "John") {
    id
    name
  }
}
        \end{lstlisting}
        \centering\small(c) GraphQL
    \end{minipage}
    \caption{Comparison of Graph Query Languages}
    \label{fig:query_examples}
\end{figure}

Cypher is a declarative language originally developed for the Neo4j platform and standardized by the openCypher project \cite{OpenCypher} since 2015. 
Gremlin is a graph traversal machine and language designed, developed, and distributed by the Apache TinkerPop project with a focus on path-like expressions \cite{Rodriguez2015}. 
GraphQL, released by Facebook in 2015, differs from the other two query languages in its original purpose. 
Initially conceived as a more flexible alternative to REST APIs, GraphQL’s core, rooted in graph algebra extended from relational algebra, enables it to query property-graph structures \cite{He2008}. 
The choice of a query language for accessing information in graph databases lacks uniformity. 
The most widely supported query language across Graph Database Management Systems (GDBMS) is Gremlin, followed by GraphQL, with Cypher ranking last \cite{Mike2022}. 
However, in terms of popularity for querying graph-based systems Cypher comes first, followed by GraphQL and Gremlin \cite{Seifer2019}. 
GraphQL also has a large community outside of the Graph Database sphere, serving as an interface for services to communicate with each other. 
These discrepancies present challenges, as certain databases support access via Gremlin, while developers may be more familiar with using technologies like GraphQL for data retrieval. 
This scenario is exemplified in the case of the open-source GDBMS JanusGraph \cite{JanusGraph} and Amazon Neptune \cite{AmazonNeptune}, a cloud solution offered by Amazon Web Services. 
Both rank among the top eleven GDBMS according to the DB-Engines Ranking as of April 2026 \cite{DBEngines}.

\begin{figure}[h]
    \centering
    \includegraphics[width=0.7\textwidth]{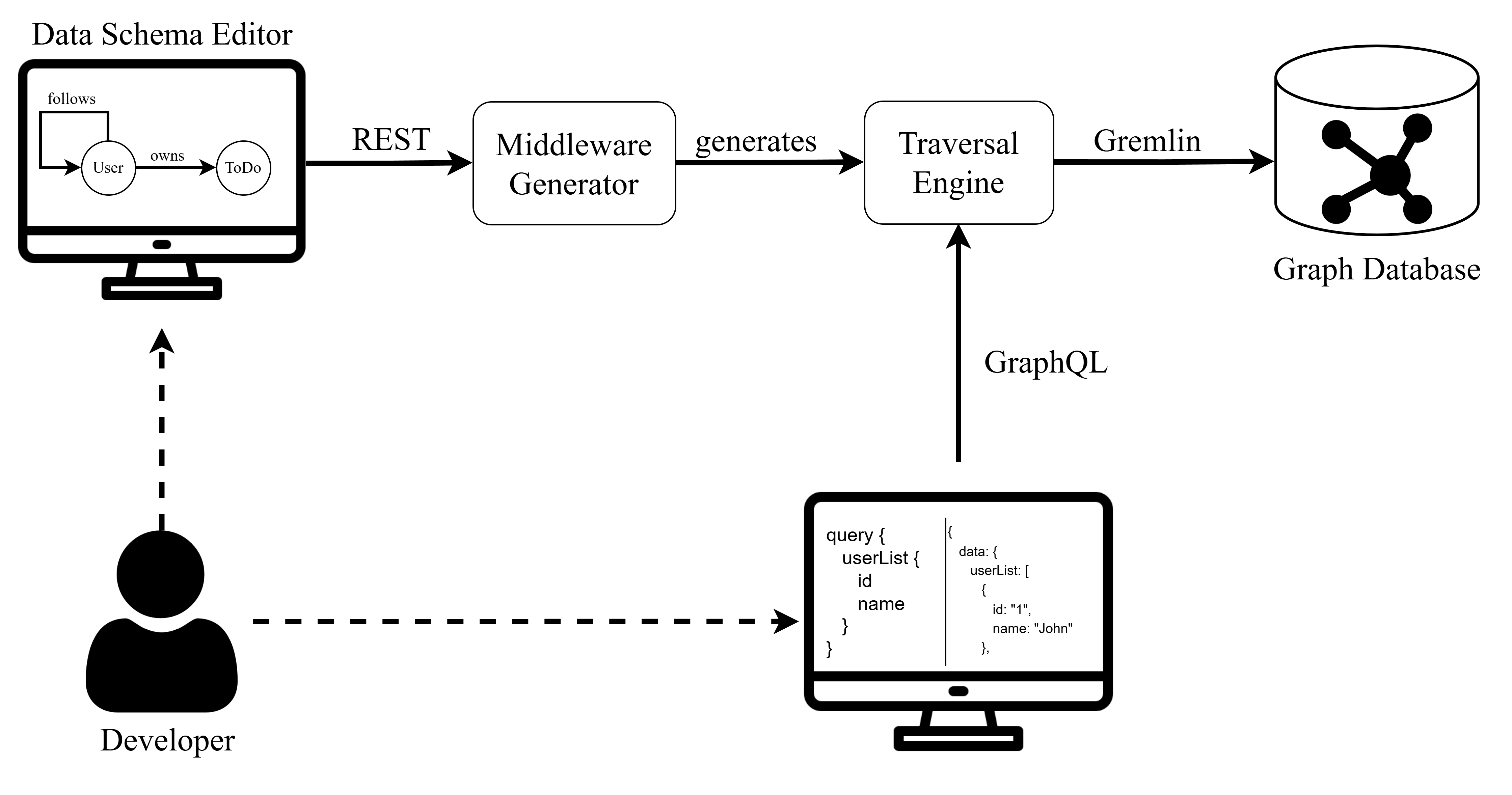}
    \caption{Architecture Overview}
    \label{fig:architecture}
\end{figure}

This paper presents Graphify, a solution aimed at resolving the challenges associated with limited schema definitions and the absence of GraphQL support for interacting with graph databases. 
The concept shown in Figure \ref{fig:architecture} empowers developers to visually construct a comprehensive data schema, encompassing property and type constraints, through a web-based Editor. 
This Editor interfaces with the Middleware Generator via a REST API to store data schemas leveraging a standardized internal data structure. 
This structure contains all requisite information to synthesize a fully functional service capable of executing complete create, read, update, delete (CRUD) operations, including nested logical predicates, multi-key sorting, and pagination, on graph databases. 
Unlike traditional middleware that often relies on manual "glue code," this approach automates the deployment of a Traversal Engine equipped to manage arbitrarily nested GraphQL requests based on the generated schema. 
The architecture of the GraphQL schema guarantees that only operations consistent with the defined data schema are permitted. 
To ensure high-performance execution, a recursive state machine algorithm processes incoming GraphQL requests, converting even complex, multi-level traversals into a single, Gremlin query. 
By leveraging information from the AST and the internal data structure, the engine effectively mitigates the $N+1$ query problem common in GraphQL-to-database interfaces. 
Eventually, the resulting Gremlin query interacts with the underlying database to retrieve or manipulate data in a type-safe manner. 

This paper primarily focuses on the processes of the Middleware Generator and Traversal Engine. 
Initially, the functionality behind the REST API of the Middleware Generator is analyzed to explore how the data schema defined by the developer is transformed into the internal data structure including the property and type constraints, and which validations are crucial to ensure a valid GraphQL schema when generating the Traversal Engine. 
Subsequently, a formal mapping is established to demonstrate how different elements of the GraphQL specification represent CRUD operations in property graph databases.
Furthermore, the process of translating GraphQL requests into Gremlin, leveraging the AST generated by the Query Parser along with the information of the internal data structure, is examined.
Finally, a formal complexity analysis establishes the linear-time efficiency of the translation, which is subsequently verified through empirical evaluation using the MovieLens 100k dataset.

\section{Related Work}
The utilization of design languages, such as the Unified Modeling Language (UML), for generating database constraints is a common practice in object-relational databases \cite{Marcos2003}. 
However, NoSQL databases, particularly graph databases, often lack support for constraints on the database level. 
To address this limitation, one approach is to ensure type safety by delegating data validation on the application layer. 
The UMLtoGraphDB tool introduces a middleware consuming an UML class diagram along with the Object Constraint Language to validate input before it reaches the graph database \cite{De2016}.
Additionally, there are proposals for syntax extensions aimed at defining constraints within the Cypher query language for the Neo4J database \cite{Pokorny2017}. 
Furthermore, efforts have been made toward achieving query language independence using transpilers. 
This involves compiling different languages into a common intermediate representation, which can be converted back to the desired query language \cite{Mike2022}.
Moreover, research has demonstrated the feasibility of converting a relational model to a graph model without semantic loss \cite{Wardani2014}. 
Additionally, investigations have been conducted into the potential implementation of a GraphQL interface on top of property graph databases \cite{Mattsson2020}. 
The solution presented in this paper proposes an approach to synthesize a fully functional backend by streamlining data validation and query representation. 
Unlike existing middleware, it leverages the GraphQL schema to handle validation tasks and utilizes a recursive state machine to convert ASTs into single, optimized Gremlin queries, thereby ensuring high performance without the need for additional components.

\section{Internal Data Structure}
In order to understand the data schema that is represented within the internal data structure, the fundamental components of a property graph database have to be explained first. 
A property graph, defined in \cite{Angles2018}, is composed of nodes, also known as vertices, denoted by a finite set $N$, and edges represented by a finite set $E$, which is disjoint from $N$.
The function $\rho$ associates each edge $E$ with a pair of nodes, thereby indicating the graph’s directed nature.
Additionally, the function $\lambda$ assigns a label $L$ to every node and edge in the graph, thereby categorizing them into specific groups.
Finally, the partial function $\sigma$ specifies that both nodes and edges can encompass (possibly empty) properties describing the element with a set of values. 
Property graph databases permit the insertion of an arbitrary number of nodes and edges with different labels, as well as the storage of various formations of properties within nodes or edges, even if they are classified under the same category using the label.

\subsection{Data Schema}
\begin{figure}[h]
    \centering
    \includegraphics[width=0.7\textwidth]{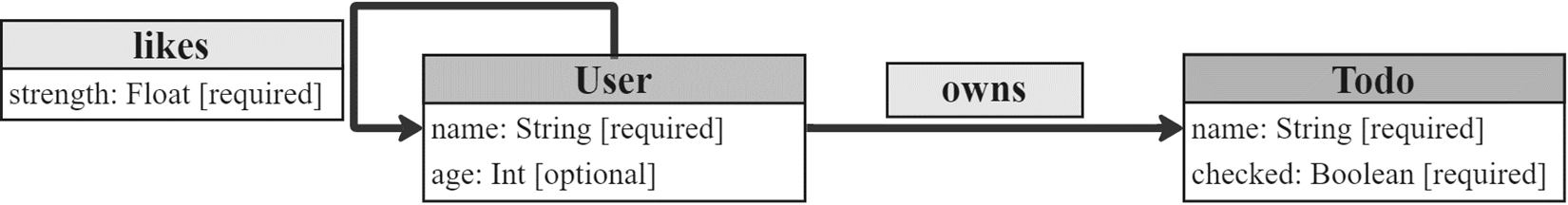}
    \caption{Todo Application Example Data Schema}
    \label{fig:data_schema}
\end{figure}

This paper presents an approach to define the data schema to model the data to be stored in the database. 
In Figure \ref{fig:data_schema}, an illustrative example demonstrates how a developer can establish a data schema for organizing a simple Todo application. 
The schema clarifies that only two categories of vertices may exist, categorized by the labels \textit{User} and \textit{Todo}. 
Arrows signify that an association can be established between nodes labeled \textit{User} through an edge labeled \textit{likes}. 
Additionally, the schema permits edges in a defined direction labeled \textit{owns} between the nodes of \textit{User} and \textit{Todo}. 
For each vertex and edge, the data schema specifies constraints to define the datatype and necessity of properties.

\subsection{Class Hierarchy}
The internal data structure has to be capable of describing a data schema created by the developer based on the artifacts of property graph databases, manifested as a collection of object-oriented structures within a programming language. 
To reconcile these concepts, a class Graph is employed to encapsulate a finite set of nodes and edges, organized as instances of the Vertex and Edge classes, respectively. 
The association of each edge with a pair of nodes is facilitated through the references to the connected vertices contained within the Edge class.
This bidirectional adjacency modeling is critical for the Traversal Engine's ability to resolve arbitrarily nested queries in a single database pass. 
Each vertex and edge instance is classified via the label field and equipped with a list of properties. 
The fields required and datatype of the Property class extend the definition of property graph databases, by specifying required fields and datatype constraints, ensuring the integrity of values stored as properties in the database. 
The types defined in the Datatype enumeration correspond to the built-in scalar types of the GraphQL specification \cite{GraphQLSpec}.

\begin{figure}[h]
    \centering
    \includegraphics[width=1\textwidth]{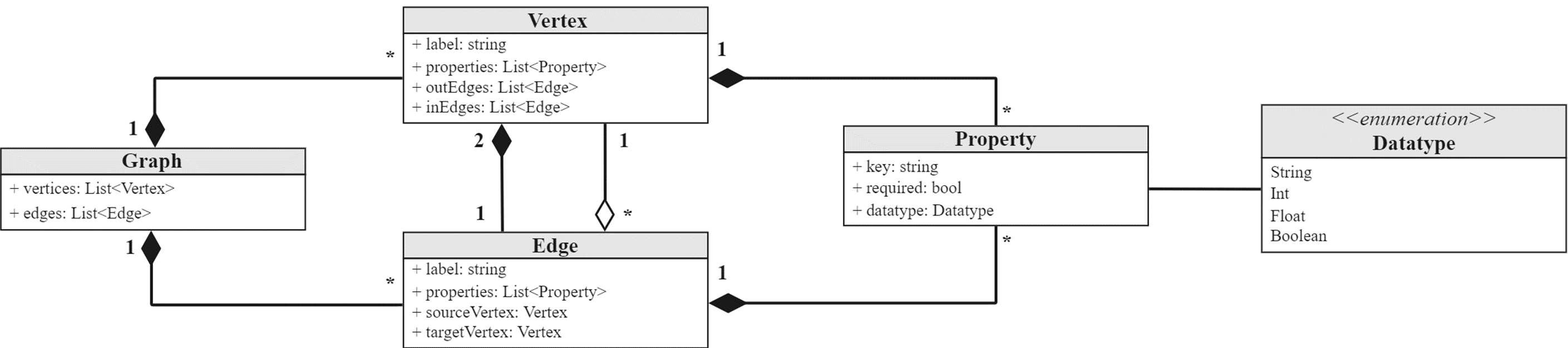}
    \caption{Internal Data Structure UML Class Diagram}
    \label{fig:uml}
\end{figure}

The structure depicted in Figure \ref{fig:uml} incorporates all essential details for the generation of the Traversal Engine. 
Each vertex maintains a list of references to both outgoing and incoming edges, while each edge preserves references to its source and target vertex. 
This arrangement enables graph traversal by navigating through the adjacent elements of each vertex or edge. 
This functionality is inevitable for representing the data model as a GraphQL schema and facilitating the process of converting GraphQL requests to Gremlin queries.

\subsection{Validations} \label{subsec:validations}
To maintain the integrity of the internal data structure and ensure the generated GraphQL schema is always valid and collision-free, adherence to specific rules is imperative. 
Upon receiving a request via the REST API, the Middleware Generator employs a set of validators to verify the following conditions:

\begin{itemize}
  \item The label of a vertex/edge must start with a letter or underscore, followed by letters, numbers, or underscores
  \item The label of a vertex/edge must not start with two underscores
  \item The key of a property must start with a letter or underscore, followed by letters, numbers, or underscores
  \item The key of a property must not be id, label, or start with two underscores
  \item The key of a property must be unique inside the vertex/edge specified
  \item The label of a vertex must be unique
  \item An edge must be connected to a source and a target vertex
  \item The label of a vertex must differ from the property keys of the connected edges
  \item The label of an edge must be unique among the edges pointing in the same direction of the connected vertices
  \item The label of an edge appended with \textit{In} must differ from the property keys of the target vertex
  \item The label of an edge appended with \textit{Out} must differ from the property keys of the source vertex
\end{itemize}

\section{Formal Mapping of GraphQL to Property Graph Semantics} \label{sec:mapping}
This chapter elucidates how a GraphQL schema is structured to include CRUD operations based on the data schema defined by the developer. 
This schema enforcement guarantees that only operations consistent with the predefined data schema are authorized. 
GraphQL delineates data retrieval and modification concerns through constructs known as \textit{Query} and \textit{Mutation} \cite{GraphQLQueries}. 
The ensuing section delves into the formal explanation of how the artifacts of property graph databases can be mapped to a GraphQL schema for reading operations. 
Additionally, certain operations are shown to filter the returned data of a GraphQL request via arguments. 
The subsequent section explains the representation of data manipulation operations within a GraphQL schema. 
As the GraphQL schema intricately hinges upon the developer-defined data schema, the Middleware Generator is tasked with generating the entire GraphQL schema based on the internal data structure.

\subsection{Query}
The connection between the fundamental components of property graph databases and the elements of a GraphQL schema is explored through an examination of a formal GraphQL schema definition \cite{ApolloSchema, GraphQLSpec}.
Therefore, three infinite countable disjunct sets consisting of $\mathit{Fields}$, $\mathit{Arguments}$, and $\mathit{Types}$ are considered.
Additionally, there exists a finite set of $\mathit{Scalars} \subset \mathit{Types}$ with a set of scalar values, referred to as $\mathit{Vals}$, and a function $\mathit{values} \colon \mathit{Scalars} \to 2^{\mathit{Vals}}$ which assigns a set of values to every scalar type.
The finite subsets $F \subset \mathit{Fields}$, $A \subset \mathit{Arguments}$, and $T \subset \mathit{Types}$ of the above sets are utilized to define GraphQL schemas.
$T$ is represented by the disjoint union of $\mathit{O_T}$ (object types), $\mathit{I_T}$ (interface types), and $\text{Scalars}$.
Lastly, $\mathit{L_T}$ gets denoted as the set $\{[t] \mid t \in T\}$ with $[t]$ representing list types.
Although the formal model is simplified for presentation, the underlying transpiler implementation encompasses the complete GraphQL specification, including artifacts such as input types, enums, and inheritance. This section maintains focus exclusively on the fundamental mapping of object types to graph semantics to ensure theoretical clarity.

Consequently, a GraphQL schema $S$ over $(F, A, T)$, representing the necessary components for data retrieval in property graph databases, comprises the following rules as outlined in Definition 2.1 in \cite{ApolloSchema}.
Each assignment is elucidated with parallels to the artifacts of property graph databases:

\begin{itemize}
  \item $\mathit{fields}_S : (\mathit{O_T} \cup \mathit{I_T}) \to 2^F$ maps every object and interface type to a set of fields. This facilitates the representation of nodes and edges as $\mathit{O_T}$ with a set of properties as $\mathit{fields}_S$.
  \item $\mathit{args}_S : F \to 2^A$ assigns a group of arguments to each field, suitable to define operations for filtering the results getting returned by a query.
  \item $\mathit{type}_S : F \cup A \to T \cup \mathit{L_T}$ assigns to each field and argument a type or list type, facilitating access to adjacent elements in the property graph through fields of $\mathit{O_T}$ representing nodes or edges.
  \item $\mathit{implementation}_S : \mathit{I_T} \to 2^{\mathit{O_T}}$ maps interface types to sets containing object types, with every $\mathit{O_T}$ implementing at least the fields of $\mathit{I_T}$. Interfaces with fields comprising id and label can generalize each $\mathit{O_T}$ representing a vertex or edge, as the graph database generates an identifier and attaches a label to them by definition.
\end{itemize}

The data schema of the Todo application shown in Figure \ref{fig:data_schema} gives the following GraphQL schema $S$ over $(F, A, T)$.
The set $\mathit{Fields}$ comprises three disjunct sets $F_D$, $F_P$, and $F_A$, denoted as $F = F_D \cup F_P \cup F_A$.
The set $F_D$ represents the default fields id and label, while $F_P$ encompasses all property keys defined in both nodes and edges.
Additionally, $F_A$ includes the lowercase labels of edges appended with In and Out along with the lowercase names of the vertices, facilitating navigation to adjacent elements in the property graph through fields.
For simplicity, the arguments $A$ are temporarily omitted and will be specified separately later in this section.
The set of \textit{Types}, as per formalization, is defined as $T = \mathit{O_T} \cup \mathit{I_T} \cup \mathit{Scalars}$, with all vertices and edges (one edge object type for each combination of source and target vertex) represented as $\mathit{O_T}$.
The elements of the sets adhere to the GraphQL naming conventions \cite{ApolloSchema}.
To prevent schema violations through duplicate names, a Vertex or Edge suffix is appended to the corresponding object type.
Furthermore, the rules specified in section \ref{subsec:validations} ensure the disjointness of the subsets of $F$.
\begin{gather*}
\begin{aligned}
    F_D &= \{id, label\} \\
    F_P &= \{name, age, checked, strength\} \\
    F_A &= \{likesIn, likesOut, ownsIn, ownsOut, todo, user\} \\
    \mathit{O_T} &= \{\mathrm{UserVertex, TodoVertex, UserToUserLikesEdge, UserToTodoOwnsEdge,} \\
    \quad &\quad \quad \mathrm{TodoToUserOwnsEdge}\} \\
    \mathit{I_T} &= \{\mathrm{GraphElement}\} \\
    \mathit{Scalars} &= \{\mathrm{ID, String, Int, Float, Boolean}\}
\end{aligned}
\end{gather*}
By definition, a property graph database organizes its elements, comprising nodes and edges, in alternating order. 
Consequently, vertices cannot be connected without an edge, and conversely, edges cannot exist without vertices. 
To facilitate traversal to adjacent elements within the GraphQL schema, node object types encompass all connected edges as fields, while edge object types require a field referencing the vertex to which the edge points. 
Considering this, the exemplary $\mathit{fields}_S$ for the interface type \textit{GraphElement}, the node \textit{User} and the edge \textit{likes} ensue:
\begin{gather*}
\begin{aligned}
    \text{GraphElement} &\to \{id, label\} \\
    \text{UserVertex} &\to \{id, label, name, age, likesIn, likesOut, ownsOut\} \\
    \text{UserToUserLikesEdge} &\to \{id, label, strength, user\}
\end{aligned}
\end{gather*}
According to the property datatypes and the cohesion of the different elements specified in the data schema, the function $\mathit{type}_S$ , restricted for simplicity to the node \textit{User} and the edge \textit{likes}, is defined as follows:
\begin{gather*}
\begin{aligned}
    \text{id}         &\rightarrow \text{ID} \\
    \text{label}      &\rightarrow \text{String} \\
    \text{name}       &\rightarrow \text{String} \\
    \text{age}        &\rightarrow \text{Int} \\
    \text{strength}   &\rightarrow \text{Float} \\
    \text{likesIn}    &\rightarrow [\text{UserToUserLikesEdge}] \\
    \text{likesOut}   &\rightarrow [\text{UserToUserLikesEdge}] \\
    \text{ownsOut}    &\rightarrow [\text{UserToTodoOwnsEdge}] \\
    \text{user}       &\rightarrow \text{UserVertex}
\end{aligned}
\end{gather*}
Given that the \textit{likes} edge, embodied by the object type \textit{UserToUserLikesEdge}, consistently points to the same vertex regardless of its orientation, defining the type in the set $\mathit{O_T}$ is necessary only once. 
Conversely, the \textit{owns} edge must be included twice in the set $\mathit{O_T}$ to account for the varying target vertex and, subsequently, the varying type of the field based on the edge’s direction. 
Besides that, the $\mathit{implementation}_S$ for the interface type, illustrated by the node User and the edge likes, is provided below:
\begin{gather*}
\begin{aligned}
    \text{implementations}_S &= \{ \mathrm{UserVertex, UserToUserLikesEdge} \} 
\end{aligned}
\end{gather*}
To define the GraphQL schema, \textit{S} considers an additional distinguished type called the root type, formally denoted as $\mathit{root}_S \subset \mathit{O_T}$ \cite{ApolloSchema}. 
These types serve as entry points for querying information through GraphQL of the underlying data source. 
In the context of property graph databases, these root types correspond to the object types derived from nodes in the initial data schema. 
Each of these root types can be directed to yield a single element by providing an additional identifier as an argument or to retrieve a list type $\mathit{L_T}$ of vertices matching the return type. 
Consequently, the set $\mathit{root}_S$ exemplary for the node \textit{User} is defined as follows:
\begin{gather*}
\begin{aligned}
    \mathit{root}_S &= \{ \mathrm{UserVertex}, [\mathrm{UserVertex}] \} 
\end{aligned}
\end{gather*}
On the left-hand side of Figure \ref{fig:schema_query_result} an excerpt of the actual GraphQL schema, derived from the formal definition, is displayed. 
For simplicity, all arguments except \textit{id} in \textit{Query} are removed. 
Next to the schema, a GraphQL query is shown to request the \textit{strength} of the \textit{likes} relationships together with the \textit{name} of all users connected to the user with an \textit{id} of 1. 
An exemplary result of the conducted request is displayed on the right-hand side:

\begin{figure}[H]
    \centering
    \begin{minipage}[b]{0.45\textwidth}
        \begin{lstlisting}[language=SQL, basicstyle=\small\ttfamily, columns=flexible]
type Query {
  user(id: ID!): UserVertex!
  userList: [UserVertex!]!
}
type UserVertex implements GraphElement {
  id: ID!
  label: String!
  name: String!
  age: Int!
  likesIn: [UserToUserLikesEdge!]!
  likesOut: [UserToUserLikesEdge!]!
  ownsOut: [UserToTodoOwnsEdge!]!
}
        \end{lstlisting}
        \centering\small(a) GraphQL Schema
    \end{minipage}
    \hfill
    \begin{minipage}[b]{0.2\textwidth}
        \begin{lstlisting}[language=Java, basicstyle=\small\ttfamily]
query {
  user(id: "1") {
    likesOut {
      strength
      user {
        name
      }
    }
  }
}
        \end{lstlisting}
        \centering\small(b) GraphQL Query
    \end{minipage}
    \hfill
    \begin{minipage}[b]{0.3\textwidth}
        \begin{lstlisting}[language=HTML, basicstyle=\small\ttfamily]
"data": {
  "user": {
    "likesOut": [
      {
        "strength": 0.73,
        "user": {
          "name": "Bob"
        }
      }
    ]
  }
}
        \end{lstlisting}
        \centering\small(c) GraphQL Result
    \end{minipage}
    \caption{Basic Data Retrieval Workflow}
    \label{fig:schema_query_result}
\end{figure}

The structure of arranging nodes and edges within the GraphQL schema allows the expression of intricate patterns leveraging the GraphQL query language. 
This enables the developer to navigate in virtually any direction inside the underlying property graph. 
Moreover, the GraphQL schema serves the purpose of documentation of the stored data model, enabling the developer for type-safe interactions with the database.

In addition to defining nodes and edges, the GraphQL schema generated by the Middleware Generator includes operations for manipulating the data returned by queries. 
These operations enable filtering based on conditions, defining the order of elements, and limiting the number of returned items. 
As these operators can be applied to both vertices and edges, the arguments \textit{A} differentiate between these contexts:
\begin{gather*}
\begin{aligned}
    A &= \{ \mathrm{where, whereVertex, whereEdge, orderBy, orderByVertex, orderByEdge, pagination} \} 
\end{aligned}
\end{gather*}
The named operations define techniques to manipulate the output of a list. 
Therefore, the arguments are applied to the fields returning a list type $\mathit{L_T}$. 
As a result, each field that returns a list of edges in an object type, identified as a node, is associated with specific operations. 
The argument names are partially appended with \textit{Vertex} or \textit{Edge} to determine whether the operation should apply to the upcoming edges or the subsequent vertex. 
For root types returning a list of vertices rather than a list of edges with a subsequent vertex, this distinction is not necessary. 
Hence, the operations are named more simply to maintain readability. 
As an example, the function $\mathit{args}_S$ is defined for both the user root type and object type as follows:
\begin{gather*}
\begin{aligned}
    \mathrm{userList} &\rightarrow \{ \mathrm{where, orderBy, pagination} \} \\
    \mathrm{likesIn}  &\rightarrow \{ \mathrm{whereVertex, whereEdge, orderByVertex, orderByEdge, pagination} \} \\
    \mathrm{likesOut} &\rightarrow \{ \mathrm{whereVertex, whereEdge, orderByVertex, orderByEdge, pagination} \} \\
    \mathrm{ownsOut}  &\rightarrow \{ \mathrm{whereVertex, whereEdge, orderByVertex, orderByEdge, pagination} \}
\end{aligned}
\end{gather*}
The utilized arguments closely adhere to established solutions in existing GDBMS such as Neo4J’s GraphQL library \cite{Neo4jGraphQL}, ensuring consistency. 
In Figure \ref{fig:query_arguments}, the GraphQL schema defines the root type of the user node, returning a list type and therefore specifying arguments. 
Notably, the \textit{UserVertex} type encompasses arguments for every field returning a list type of edges. 
However, for simplicity, the object type is abbreviated to display only the \textit{likesOut} field, which includes all available arguments. 
The query displayed next to the schema demonstrates the utilization of some arguments to retrieve for each user named \textit{John}, the two users older than 18 whom he likes the most. 
Assigning a dedicated input object \cite{GraphQLSpec} for each argument enables type-safe operations by predefining filter operations and storing properties of both nodes and edges as an enum type \cite{GraphQLSpec}.

\begin{figure}[ht]
    \centering
    \begin{minipage}[b]{0.48\textwidth}
        \begin{lstlisting}[
            language=SQL, 
            basicstyle=\footnotesize\ttfamily, 
            columns=flexible, 
            tabsize=2,
            breaklines=true
        ]
type Query {
  userList(
    where: UserVertexLogicInput
    orderBy: [UserVertexOrderByInput!]
    pagination: PaginationInput
  ): [UserVertex!]!
}
type UserVertex implements GraphElement {
  likesOut(
    whereVertex: UserVertexLogicInput
    orderByVertex: [UserVertexOrderByInput!]
    whereEdge: UserToUserLikesEdgeLogicInput
    orderByEdge: [UserToUserLikesEdgeOrderByInput!]
    pagination: PaginationInput
  ): [UserToUserLikesEdge!]!
}
        \end{lstlisting}
        \centering\footnotesize (a) GraphQL Schema
    \end{minipage}
    \hfill
    \begin{minipage}[b]{0.48\textwidth}
        \begin{lstlisting}[
            language=Java, 
            basicstyle=\footnotesize\ttfamily, 
            columns=flexible, 
            tabsize=2,
            breaklines=true
        ]
query {
  userList(where: {name_EQ: "John"}) {
    likesOut (
      pagination: {offset: 0, limit: 2}
      orderByEdge: {
        property: strength, 
        order: DESC
      }
      whereVertex: {age_GT: 18}
    ) {
      user {
        name
      }
    }
  }
}
        \end{lstlisting}
        \centering\footnotesize (b) GraphQL Query
    \end{minipage}
    \caption{Data Retrieval with Arguments}
    \label{fig:query_arguments}
\end{figure}

\subsection{Mutation}
For creating and updating nodes, an input type is specified, encompassing all properties with the datatype specified in the data schema as fields. 
In case the property is denoted as required in the data schema, the corresponding datatype within the input type of the GraphQL schema is marked with an exclamation mark. 
This approach ensures that developers adhere to the rules of the defined data schema when inserting and updating data, thereby ensuring data integrity within the database. 
Property graph databases automatically generate a unique identifier for every stored node. 
This identifier is returned upon the creation of a node and is utilized in update operations to identify the specific node being modified. 
Below in Figure \ref{fig:add_update_mutation} is an excerpt of the GraphQL schema, showcasing the root types for creating and updating nodes of a user through an input type, along with corresponding GraphQL requests:

\begin{figure}[H]
    \centering
    \begin{minipage}[b]{0.48\textwidth}
        \begin{lstlisting}[
            language=SQL,
            basicstyle=\footnotesize\ttfamily,
            columns=flexible,
            tabsize=2,
            breaklines=true
        ]
type Mutation {
  addUserVertex(data: UserVertexInput!): ID!
  updateUserVertex(id: ID!, data: UserVertexInput!): ID!
}
input UserVertexInput {
  name: String!
  age: Int
}
        \end{lstlisting}
        \centering\footnotesize (a) GraphQL Schema Vertex Create and Update
    \end{minipage}
    \hfill
    \begin{minipage}[b]{0.48\textwidth}
        \begin{lstlisting}[
            language=Java, 
            basicstyle=\footnotesize\ttfamily, 
            columns=flexible, 
            tabsize=2,
            breaklines=true
        ]
mutation {
  addUserVertex(data: {name: "John"})
}
mutation {
  updateUserVertex(data: {name: "Bob", age: 7})
}
        \end{lstlisting}
        \centering\footnotesize (b) GraphQL Mutation Vertex Create and Update
    \end{minipage}
    \caption{Vertex Mutation with Add and Update Operations}
    \label{fig:add_update_mutation}
\end{figure}

Analogous to the nodes, edges are defined with an input type based on the properties specified in the data schema as well. 
The update operation mirrors that of nodes utilizing a unique identifier to reference the corresponding edge. 
Connecting two nodes via an edge requires the identifiers of the source and target vertices to ensure their existence. 
The naming convention of the root types facilitates the use of edges with the same label between different nodes. 
The following example in Figure \ref{fig:add_update_mutation_edge} illustrates the root types along with operations for the \textit{likes} edge.

\begin{figure}[H]
    \centering
    \begin{minipage}[b]{0.48\textwidth}
        \begin{lstlisting}[
            language=SQL,
            basicstyle=\footnotesize\ttfamily,
            columns=flexible,
            tabsize=2,
            breaklines=true
        ]
type Mutation {
  connectUserToUserViaLikesEdge(
    source_user_id:ID!, 
    target_user_id: ID!
    data: UserToUserViaLikesEdgeInput!
  ): ID!
  updateUserToUserViaLikesEdge(
    id: ID!, 
    data: UserToUserViaLikesEdgeInput!
  ): ID! 
}
input UserToUserViaLikesEdgeInput {
  strength:Float!
}
        \end{lstlisting}
        \centering\footnotesize (a) GraphQL Schema Edge Create and Update
    \end{minipage}
    \hfill
    \begin{minipage}[b]{0.48\textwidth}
        \begin{lstlisting}[
            language=Java, 
            basicstyle=\footnotesize\ttfamily, 
            columns=flexible, 
            tabsize=2,
            breaklines=true
        ]
mutation {
  connectUserToUserViaLikesEdge (
    source_user_id: "1",
    target_user_id: "2",
    data: {strength: 0.73}
  )
}
mutation {
  updateUserToUserLikesEdge (
    data: {strength: 0.37}
  )
}
        \end{lstlisting}
        \centering\footnotesize (b) GraphQL Mutation Edge Create and Update
    \end{minipage}
    \caption{Edge Mutation with Add and Update Operations}
    \label{fig:add_update_mutation_edge}
\end{figure}

Since deletion requires only the identifier generated by the database for both nodes and edges, no additional information is necessary. 
Therefore, there is no need to specify individual root types with dedicated naming conventions for each node/edge. 
The delete root types within the mutation section of the GraphQL schema depicted in Figure \ref{fig:detete} are sufficient to remove every vertex and edge existing in the database.

\begin{figure}[H]
    \centering
    \begin{minipage}[b]{0.48\textwidth}
        \begin{lstlisting}[
            language=SQL,
            basicstyle=\footnotesize\ttfamily,
            columns=flexible,
            tabsize=2,
            breaklines=true
        ]
type Mutation {
    deleteVertex(id: ID!): ID!
    deleteEdge(id: ID!): ID!
}
        \end{lstlisting}
        \centering\footnotesize (a) GraphQL Schema Vertex and Edge Delete
    \end{minipage}
    \hfill
    \begin{minipage}[b]{0.48\textwidth}
        \begin{lstlisting}[
            language=Java, 
            basicstyle=\footnotesize\ttfamily, 
            columns=flexible, 
            tabsize=2,
            breaklines=true
        ]
mutation {
  deleteVertex(id: "1")
}
mutation {
  deleteEdge(id: "3")
}
        \end{lstlisting}
        \centering\footnotesize (b) GraphQL Query Vertex and Edge Delete
    \end{minipage}
    \caption{Vertex and Edge Delete Operations}
    \label{fig:detete}
\end{figure}

\section{Conversion of GraphQL Requests to Gremlin Queries}
\label{sec:conversion}
To interact with the database, the Traversal Engine must convert incoming GraphQL requests into Gremlin queries. 
Therefore, the Middleware Generator generates resolvers for each root type along with the GraphQL schema. 
When a GraphQL request is received, first the query is parsed by the utilized GraphQL framework into an AST, which is then validated against the schema. 
If the query is valid, the corresponding resolver function is invoked, with the AST passed as a parameter. 
Alongside the data structure, which is initially conveyed to the Traversal Engine as well, the AST contains the necessary information to generate the corresponding Gremlin query. 
While the mutation resolvers just insert the parameters passed into a predefined Gremlin query, the query resolvers require a more sophisticated approach to cover all possible traversals through the property graph. 
The following sections describe the concept of converting the AST into a single Gremlin query through a generic algorithm.

\subsection{Gremlin}
Formally, Gremlin, as a graph traversal machine, comprises a graph $G$, a traversal $\Psi$, and a set of traversers $T$.
The traversal $\Psi$ represents a tree of functions, either arranged as a linear chain $f \circ g \circ h$ or nested $f(g \circ h) \circ k$ \cite{Rodriguez2015}. 
This nested motif combined with the project step \cite{GremlinProjectStep} of the Gremlin query language, constitutes the basic building blocks of the conversion mechanism. 
The function-chaining behavior of Gremlin can be seamlessly represented in various programming languages \cite{TinkerPopGithub} as a set of chained functions. 
The example depicted in Figure \ref{fig:gremlin_conversion} demonstrates a GraphQL request converted into Gremlin, implemented in Python \cite{Python} utilizing the gremlinpython library \cite{GremlinPython}.

\begin{figure}[H]
    \centering
    \noindent 
    \begin{minipage}[b]{0.30\textwidth}
        \begin{lstlisting}[
            language=SQL,
            basicstyle=\footnotesize\ttfamily,
            columns=flexible,
            tabsize=2,
            breaklines=true
        ]
query {
  userList(where: {age_GT: 18}) {
    age
    likesOut(pagination: {
      offset: 0, 
      limit: 3}
    ) {
      strength
      user {
        id
      }
    }
  }
}
        \end{lstlisting}
        \centering\footnotesize (a) GraphQL Request
    \end{minipage}
    \hfill 
    \begin{minipage}[b]{0.68\textwidth}
        \begin{lstlisting}[
            language=Java, 
            basicstyle=\footnotesize\ttfamily, 
            columns=flexible, 
            tabsize=2,
            breaklines=true
        ]
g.V().has_label('User').has('age', P.gt(18)) \
  .project('age', 'likesOut') \
  .by(__.coalesce(__.values('age'), __.constant(None))) \
  .by(
    __.out_e('likes').project('strength', 'user')
      .by(__.values('strength'))
      .by(
        __.in_v().has_label('User').project('id')
          .by(__.id_())
      ).skip(0).limit(3).fold()
  ).to_list()
        \end{lstlisting}
        \centering\footnotesize (b) Gremlin Conversion in Python
    \end{minipage}
    \caption{Gremlin Conversion of a GraphQL Query}
    \label{fig:gremlin_conversion}
\end{figure}

\subsection{Conversion State Machine}
Figure \ref{fig:gremlin_conversion} demonstrates the similarities between the two languages, illustrating how the fields of the GraphQL request are represented as a project step in Gremlin. 
Each level of hierarchy results in a nested function that includes another projection incorporating fields with nested functions once again. 
This recurring pattern can be conceptualized as a state machine, gradually chaining functions of the traversal $\Psi$ based on the AST until the query is completed. 
The state machine relies on indirect recursion to facilitate the conversion of arbitrary deeply nested GraphQL requests. 
Hence, the state machine depicted below in Figure \ref{fig:statemachine} comprises several nodes, representing functions calling one another.

\begin{figure}[H]
    \centering
    \includegraphics[width=0.45\textwidth]{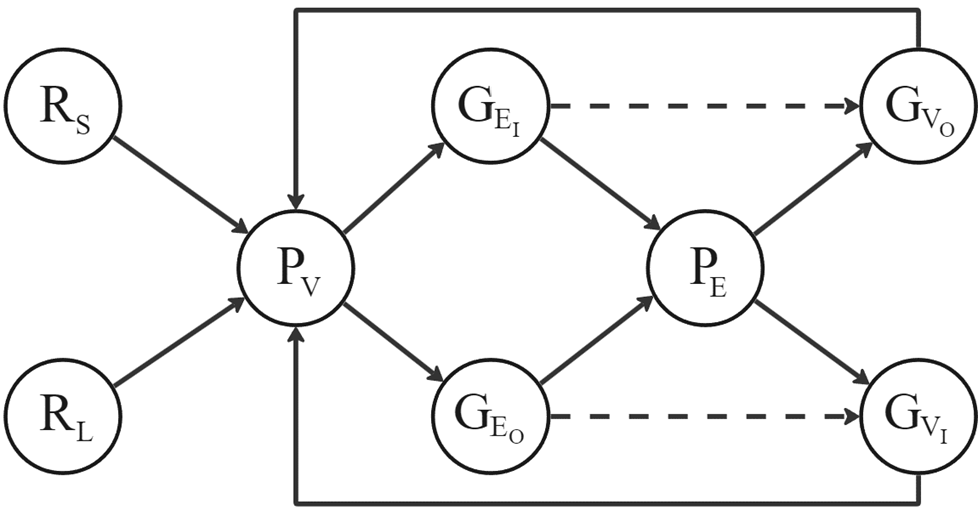}
    \caption{Conversion State Machine}
    \label{fig:statemachine}
\end{figure}

Two stack data structures containing vertices and edges, along with the internal data structure, are utilized as a reference to determine the actions to perform in the corresponding states of the machine. 
Each state, pushing a node or edge onto one of the stacks, also pops the element after the chain of consecutive functions represented by the states finishes. 
The states can be classified into three different categories serving different purposes.
The two root states depicted on the left-hand side of the state machine in Figure \ref{fig:statemachine} serve as entry points and are therefore directly connected to the generated resolver functions. 
$R_S$ is invoked if a single element is queried, while $R_L$ is responsible for list types. 
These states are responsible for preprocessing the AST, pushing the requested node onto the vertex stack, and applying operations depending on the arguments specified for the root type.
Both root states lead to the state $P_V$, responsible for converting the specified vertex fields into a projection step. 
The same regarding edges applies to $P_E$. 
To determine which function of $\Psi$ has to be applied, the state acquires information about the field types from the vertex on top of the vertex stack. 

\begin{table}[H]
  \centering
  \begin{tabular}{ll}
    \toprule
    Field Type     & Function (pythongremlin)      \\
    \midrule
    id                 & \lstinline|__.id_()| \\
    label              & \lstinline|__.label()| \\
    required field     & \lstinline|__.values(<field>)| \\
    optional field     & \lstinline|__.coalesce(__.values(<field>), __.constant(None))| \\
    \bottomrule
  \end{tabular}
  \caption{Projection State Functions}
  \label{tab:projection_table}
\end{table}

If the field falls within a classification not enumerated in the preceding table, it either references an edge $P_V$, or a vertex in the case of $P_E$. 
Under this condition, the corresponding state \textit{G} for generating a nested traversal is entered. 
Depending on whether the field represents an outgoing or incoming edge, $P_V$ navigates to the corresponding $G_E$. 
If the edge includes a field pointing to a node, $P_E$ further navigates to the $G_V$ states. 
The dashed line indicates the $G_V$ state to be invoked, which is determined by the preceding $G_E$ states. 
In addition to applying specified arguments and pushing the node/edge defined by the field onto the stack for the upcoming \textit{P} state, a nested traversal is generated. 
Table \ref{tab:conversion_table} lists the states \textit{G} with the corresponding functions used to generate the traversal.

\begin{table}[H]
  \centering
  \begin{tabular}{llll}
    \toprule
    $G_E$ State & Function (gremlinpython) & $G_V$ State & Function (gremlinpython) \\
    \midrule
    $G_{E_I}$ & \lstinline|__.in_e()|  & $G_{V_I}$ & \lstinline|__.in_v()| \\
    $G_{E_O}$ & \lstinline|__.out_e()| & $G_{V_O}$ & \lstinline|__.out_v()| \\
    \bottomrule
  \end{tabular}
 \caption{Traversal Generation State Functions}
  \label{tab:conversion_table}
\end{table}

\section{Evaluation}

The translation algorithm described in Section~\ref{sec:conversion} processes each node of the GraphQL AST exactly once through the recursive state machine shown in Figure~\ref{fig:statemachine}.
Prior to empirical benchmarking, this section establishes the formal complexity to provide a theoretical upper bound on the transpilation process.

\subsection{Complexity Analysis}
\label{sec:complexity}

Table~\ref{tab:notation} introduces the symbols used throughout this analysis.
The schema-level constants $P_{\max}$ and $\delta_{\max}$ are fixed for a given schema and treated as constants in all complexity expressions.

\begin{table}[H]
  \centering
  \begin{tabular}{cl}
    \toprule
    Symbol & Meaning \\
    \midrule
    $S$              & Total selected fields across all nesting levels (after fragment expansion) \\
    $D$              & Maximum nesting depth; number of alternating vertex-edge-vertex hops \\
    $W$              & Total filter conditions across all levels \\
    $K$              & Total ordering terms across all levels \\
    $P_{\max}$       & Maximum number of properties on any vertex or edge type (schema constant) \\
    $\delta_{\max}$  & Maximum degree of any vertex type in the schema (schema constant) \\
    \bottomrule
  \end{tabular}
  \caption{Notation for Complexity Analysis}
  \label{tab:notation}
\end{table}

\subsubsection*{Time Complexity}

The translation consists of three logically distinct steps.
First, fragment and variable expansion recurses over the query selection tree, visiting each node exactly once with $O(1)$ work per node, contributing $O(S)$.
Second, filter and ordering arguments are parsed by recursively visiting every node in the corresponding condition tree exactly once, contributing $O(W + K)$ in total across all nesting levels.
Third, and most critically, the state machine performs a recursive descent over the expanded selection tree to produce the Gremlin traversal $\Psi$ (cf.\ Figure~\ref{fig:statemachine}).
The fundamental invariant of this descent is that every field across all nesting levels is visited exactly once, there is no backtracking or re-traversal.
At each field, classifying it as a scalar property or an adjacent vertex or edge requires a lookup against the schema, which costs $O(P_{\max} + \delta_{\max})$ per field and $O(S \cdot (P_{\max} + \delta_{\max}))$ in total.
Summing all three steps yields:
\begin{equation}
  T_{\text{translation}} = O\!\left(S \cdot (P_{\max} + \delta_{\max}) + W + K\right)
  \label{eq:full}
\end{equation}
Since $P_{\max}$ and $\delta_{\max}$ are schema constants, and $W, K \leq S$ (one cannot have more filter conditions or ordering terms than selected fields), Equation~\eqref{eq:full} reduces to the central result:

\begin{definition}[Translation Time Complexity]
  Let $S$ be the total number of selected fields in a GraphQL query after fragment expansion. The time complexity of the GraphQL-to-Gremlin translation is
  \begin{equation}
    \boxed{T_{\text{translation}} = O(S)}
    \label{eq:hero}
  \end{equation}
  Every AST node is visited exactly once; there is no polynomial blowup from nesting depth $D$ or field fan-out.
\end{definition}

It is worth noting that $D$ does not appear in the time bound.
Since the state machine processes each field sequentially at every level, the total work is the \emph{sum} over all nesting levels, not the product. Therefore depth contributes no combinatorial overhead.

\subsubsection*{Space Complexity}

\begin{table}[H]
  \centering
  \begin{tabular}{lll}
    \toprule
    Component & Space & Notes \\
    \midrule
    Schema graph (static)         & $O(V_s + E_s)$ & Loaded once at startup, shared across requests \\
    Traversal context stacks      & $O(D)$         & Only the path from root to current node is retained \\
    Filter condition trees        & $O(W)$         & Short-lived; one per nesting level \\
    Ordering structures           & $O(K)$         & \\
    Generated Gremlin traversal   & $O(S + W + K)$ & One step per field, one predicate per condition \\
    Recursion call stack          & $O(D)$         & Bounded by nesting depth \\
    \bottomrule
  \end{tabular}
  \caption{Space Complexity of the Translation}
  \label{tab:space}
\end{table}

The total space consumed per request is $O(S + W + K + D)$.
Since $S \geq D$ (at least one field must be selected per depth level), this simplifies to:
\begin{equation}
  S_{\text{translation}} = O(S + W + K)
  \label{eq:space}
\end{equation}

\subsubsection*{Summary}

\begin{table}[H]
  \centering
  \begin{tabular}{lll}
    \toprule
    Parameter & Time & Space \\
    \midrule
    Fields ($S$)              & $O(S)$                  & $O(S)$ \\
    Filter conditions ($W$)   & $O(W)$                  & $O(W)$ \\
    Ordering terms ($K$)      & $O(K)$                  & $O(K)$ \\
    Nesting depth ($D$)       & $O(1)$ per level        & $O(D)$ stack space \\
    Schema degree ($\delta_{\max}$) & Constant factor on $S$ & --- \\
    \bottomrule
  \end{tabular}
  \caption{Complexity Summary}
  \label{tab:complexity_summary}
\end{table}

The dominant cost is $O(S + W + K)$ in both time and space, with $D$ subsumed by $S$ in the time bound.
Note that this analysis covers the \emph{translation step only}. The complexity of executing the resulting Gremlin traversal against the database depends on graph-specific factors such as index availability and edge fan-out, which are independent of the translation.

\subsection{Empirical Evaluation}

\subsubsection*{Experimental Setup}

The benchmark uses the MovieLens~100K dataset \cite{harper2015movielens}, which comprises 100,000 ratings by 943 users across 1,682 movies, each associated with genres and user occupations.
The dataset was loaded into a JanusGraph instance, with the graph schema generated by Graphify.
The schema defines four vertex types: \textit{UserVertex}, \textit{MovieVertex}, \textit{GenreVertex}, and \textit{OccupationVertex}, connected by three edge types: \textit{rated} (User $\to$ Movie, carrying rating and timestamp), \textit{hasGenre} (Movie $\to$ Genre), and \textit{worksAs} (User $\to$ Occupation).
Each vertex and edge type exposes the full set of CRUD operations and filter, ordering, and pagination arguments as described in Section~\ref{sec:mapping}.
The complete generated GraphQL schema and all four benchmark queries are listed in Appendix~\ref{app:schema} and Appendix~\ref{app:queries} respectively.

Four GraphQL queries of increasing structural complexity were benchmarked, each executed 120 times against the live database.
The first 20 executions served as a warm-up to allow JanusGraph's JIT compiler to reach steady state; timing measurements were recorded over the remaining 100 executions.
For every execution, two durations were measured independently: the \textit{Transpilation Time} (the time to translate the GraphQL query into a Gremlin traversal) and the \textit{Database Execution Time} (the time for JanusGraph to execute the resulting traversal and return results).

\subsubsection*{Benchmark Queries}

The four queries are chosen to stress different dimensions of the complexity model.
Table~\ref{tab:queries} summarises each query along with its structural parameters.

\begin{table}[H]
  \centering
  \begin{tabular}{llrrrr}
    \toprule
    Query & Description & $S$ & $W$ & $K$ & $D$ \\
    \midrule
    SimpleLookup      & Movie lookup by exact title                          & 3  & 1 & 0 & 0 \\
    ComplexFilter     & User list with compound filter, ordering, pagination & 2  & 2 & 1 & 0 \\
    UserRatings       & Single user with 1-hop rated movies                  & 5  & 0 & 0 & 1 \\
    GenreDemographics & Genre with 3-hop nested traversal                    & 10 & 1 & 0 & 3 \\
    \bottomrule
  \end{tabular}
  \caption{Benchmark queries and their structural parameters}
  \label{tab:queries}
\end{table}

\textit{SimpleLookup} retrieves a movie list filtered by exact title, selecting three fields ($S=3$, $W=1$).
\textit{ComplexFilter} queries the user list with a compound \texttt{AND} predicate on age and gender, plus ordering and pagination ($S=2$, $W=2$, $K=1$).
\textit{UserRatings} fetches a single user by ID and traverses one hop to their five most recent rated movies ($S=5$, $D=1$).
\textit{GenreDemographics} is the structurally deepest query: starting from a single genre, it traverses three alternating vertex-edge-vertex hops through movies, their ratings filtered by score, the rating users, and finally their occupations ($S=10$, $D=3$).

\subsubsection*{Results}

\begin{figure}[H]
  \centering
  \includegraphics[width=\textwidth]{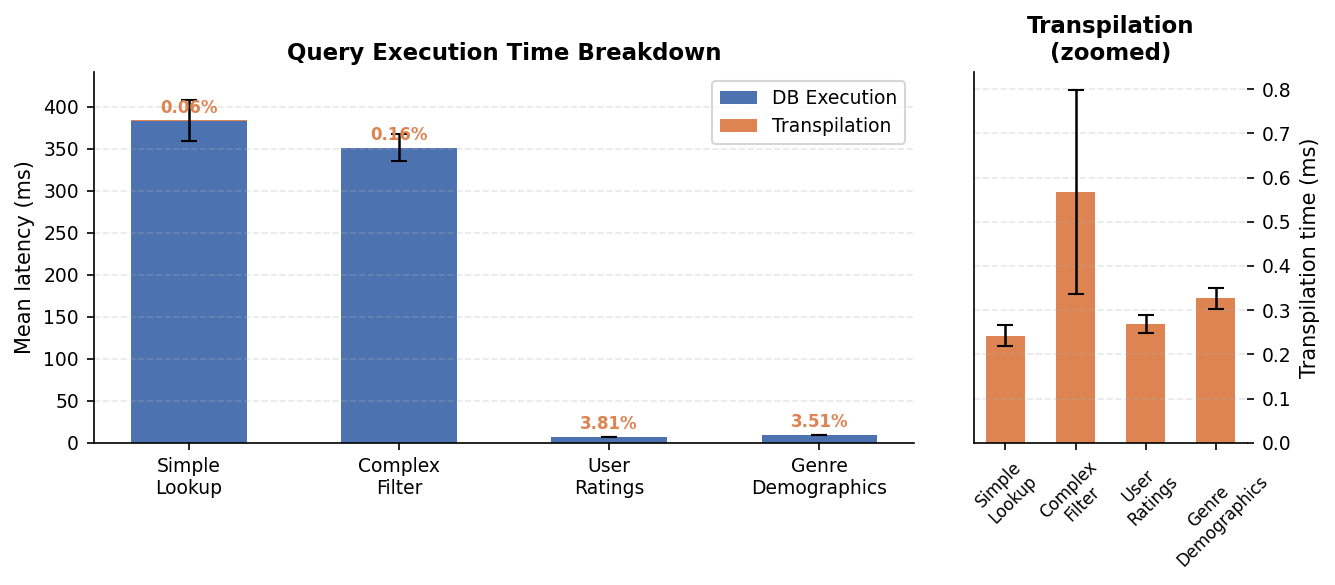}
  \caption{Query Execution Time Breakdown. The left panel shows mean DB Execution Time and Transpilation Time per query. The right panel zooms into the Transpilation Time axis to make the sub-millisecond variation visible.}
  \label{fig:evaluation}
\end{figure}

\begin{table}[H]
  \centering
  \begin{tabular}{lrrrrrrrrrr}
    \toprule
    Query & $n$ & \multicolumn{3}{c}{Transpile (ms)} & \multicolumn{3}{c}{DB Execution (ms)} & \multicolumn{2}{c}{Total (ms)} & T\% \\
    \cmidrule(lr){3-5} \cmidrule(lr){6-8} \cmidrule(lr){9-10}
     & & mean & std & p95 & mean & std & p95 & mean & p95 & \\
    \midrule
    SimpleLookup      & 102 & 0.242 & 0.123 & 0.523  & 383.2 & 124.8 & 570.2 & 383.4 & 570.5 & 0.06\% \\
    ComplexFilter     & 120 & 0.567 & 1.280 & 3.989  & 350.3 &  88.9 & 516.1 & 350.8 & 516.3 & 0.16\% \\
    UserRatings       & 120 & 0.269 & 0.110 & 0.460  &   6.8 &   2.0 &  10.1 &   7.1 &  10.3 & 3.81\% \\
    GenreDemographics & 120 & 0.327 & 0.135 & 0.685  &   9.0 &   2.4 &  13.3 &   9.3 &  13.7 & 3.51\% \\
    \bottomrule
  \end{tabular}
  \caption{Benchmark results. T\% denotes the fraction of total latency attributable to transpilation.}
  \label{tab:results}
\end{table}

\paragraph{Transpilation Time.}
Across all four queries, transpilation completes in well under one millisecond on average (0.24--0.57\,ms), consistent with the predicted $O(S)$ behaviour.
Notably, \textit{ComplexFilter} exhibits the highest mean transpilation time despite having the smallest $S=2$; this is explained by the additional argument-parsing overhead from $W=2$ filter conditions and $K=1$ ordering term, in line with Equation~\eqref{eq:full}.
The deepest query, \textit{GenreDemographics} ($S=10$, $D=3$), transpiles only marginally slower than \textit{UserRatings} ($S=5$, $D=1$), confirming that nesting depth $D$ does not independently drive translation cost.

\paragraph{Database Execution Time.}
The DB execution times diverge dramatically across queries, demonstrating that this component is governed entirely by data-graph factors rather than by $S$.
\textit{SimpleLookup} and \textit{ComplexFilter} incur the highest latency (350--383\,ms mean) because both require a full label scan over the \texttt{Movie} and \texttt{User} vertex sets respectively, as no composite index on title, age, or gender was configured.
In contrast, \textit{UserRatings} and \textit{GenreDemographics} resolve in under 10\,ms, because both queries begin with a point lookup by vertex ID (which JanusGraph serves from a primary index) and then follow a bounded number of edges via pagination.
Despite \textit{GenreDemographics} performing three nested hops against a paginated result set of considerable depth, its DB execution time (9.0\,ms) remains comparable to the shallower \textit{UserRatings} (6.8\,ms), underscoring that query depth is less consequential than index utilisation.

\paragraph{Transpilation overhead.}
As shown in the T\% column of Table~\ref{tab:results}, transpilation accounts for at most 3.81\% of total request latency across all experiments, and less than 0.2\% for the scan-heavy queries where DB execution dominates.
This confirms that the $O(S)$ transpiler introduces no measurable bottleneck relative to the cost of query execution.

\section{Conclusions and Future Work}

This paper presented Graphify, a framework that addresses two fundamental limitations of property graph databases: the absence of schema enforcement and the lack of a standardised, developer-friendly query interface.
Three concrete contributions were made.
First, a formal mapping from GraphQL schema constructs to the property graph model was established, grounding the translation in graph algebra and providing a rigorous basis for CRUD operation generation.
Second, a recursive state machine algorithm was designed and implemented that translates arbitrary GraphQL queries into a single, optimised Gremlin traversal $\Psi$.
Third, the algorithm was analysed theoretically, yielding a linear time complexity of $T_{\text{translation}} = O(S)$ in the total number of selected fields, and validated empirically on the MovieLens~100K dataset across four structurally diverse benchmark queries.

The results show that the transpilation consistently completes in well under one millisecond and accounts for at most 3.81\% of total request latency, demonstrating that the approach introduces no measurable overhead.
Database execution time, by contrast, is governed entirely by data-graph factors such as index availability and vertex fan-out, and varies by over an order of magnitude across queries, underscoring the importance of schema-aware index configuration rather than transpiler overhead as the primary performance lever.
Together, the theoretical and empirical evidence confirms that Graphify is both correct in its formal foundations and efficient in practice.

\paragraph{Future Work.}
Several directions remain open for investigation.
The most impactful near-term extension would be \textit{query-driven index recommendation}: by analysing the filter predicates and traversal patterns across incoming GraphQL queries, the system could automatically detect frequently used access paths and recommend or provision composite indices in JanusGraph, directly addressing the performance gap observed in the full label-scan queries.
A related direction is \textit{automatic query optimisation}: the transpiler currently performs a direct structural translation, but could be extended with a cost model that rewrites traversal patterns, for instance, pushing high-selectivity predicates earlier in the traversal, or reordering edge-step sequences, before emitting the final Gremlin query.
Beyond performance, \textit{schema evolution support} is a practical concern: as the data model changes, the generated GraphQL schema and resolver logic must be updated consistently, and a principled migration mechanism would reduce operational risk.
Finally, \textit{security and access control} at the GraphQL layer, field-level permissions, query depth limits, and rate limiting, would be essential for production deployments where different clients require differentiated access to the graph.

\bibliographystyle{unsrtnat}
\bibliography{references}

@inproceedings{Do2022,
  author    = {T.-T.-T. Do and T.-B. Mai-Hoang and V.-Q. Nguyen and Q.-T. Huynh},
  title     = {Query-based Performance Comparison of Graph Database and Relational Database},
  booktitle = {Association for Computing Machinery},
  address   = {Hanoi, Vietnam},
  year      = {2022}
}

@incollection{Jaroslav2015,
  author    = {P. Jaroslav},
  title     = {Graph Databases: Their Power and Limitations},
  booktitle = {Computer Information Systems and Industrial Management},
  publisher = {Springer International Publishing},
  pages     = {58--69},
  year      = {2015}
}

@inproceedings{Canovas2013,
  author    = {J. C{\'a}novas and J. Cabot},
  title     = {Discovering Implicit Schemas in JSON Data},
  booktitle = {Web Engineering},
  address   = {Aalborg, Denmark},
  publisher = {Springer Berlin Heidelberg},
  pages     = {68--83},
  year      = {2013}
}

@misc{Neo4jCypher,
  title        = {Neo4j Cypher Query Language},
  author       = {{Neo4j, Inc.}},
  year         = {2024},
  url          = {https://neo4j.com/product/cypher-graph-query-language/},
  note         = {Accessed: 2026-04-28}
}

@misc{TinkerPopGremlin,
  title        = {Apache TinkerPop™},
  author       = {{The Apache Software Foundation}},
  year         = {2023},
  url          = {https://tinkerpop.apache.org/gremlin.html},
  note         = {Accessed: 2026-04-28}
}

@misc{TinkerPopGithub,
  title        = {Apache TinkerPop GitHub Repository},
  year         = {2024},
  url          = {https://github.com/apache/tinkerpop},
  note         = {Accessed: 2026-04-28}
}

@misc{GraphQL,
  title        = {GraphQL},
  author       = {{The GraphQL Foundation}},
  year         = {2024},
  url          = {https://graphql.org/},
  note         = {Accessed: 2026-04-28}
}

@inproceedings{GraphQLSpec,
  author    = {Hartig, Olaf and Pérez, Jorge},
  title     = {Semantics and Complexity of GraphQL},
  booktitle = {Proceedings of the 2018 World Wide Web Conference (WWW '18)},
  year      = {2018},
  pages     = {1155--1164},
  publisher = {ACM},
  address   = {Lyon, France},
  doi       = {10.1145/3178876.3186014}
}

@misc{OpenCypher,
  title        = {openCypher},
  author       = {{Neo4j, Inc.}},
  year         = {2024},
  url          = {https://opencypher.org/},
  note         = {Accessed: 2026-04-28}
}

@inproceedings{Rodriguez2015,
  author    = {M. A. Rodriguez},
  title     = {The Gremlin Graph Traversal Machine and Language},
  booktitle = {Proceedings of the 15th Symposium on Database Programming Languages},
  pages     = {1--10},
  year      = {2015}
}

@inproceedings{He2008,
  author    = {H. He and A. K. Singh},
  title     = {Graphs-at-a-time: Query Language and Access Methods for Graph Databases},
  booktitle = {Association for Computing Machinery},
  address   = {New York, NY, USA},
  year      = {2008}
}

@misc{Mike2022,
  author       = {M. Mike},
  title        = {Translating Between Graph Database Query Languages},
  institution  = {ETH Zurich},
  year         = {2022}
}

@inproceedings{Seifer2019,
  author    = {P. Seifer and J. H{\"a}rtel and M. Leinberger and R. L{\"a}mmel and S. Staab},
  title     = {Empirical Study on the Usage of Graph Query Languages in Open Source Java Projects},
  booktitle = {Proceedings of the 12th ACM SIGPLAN International Conference on Software Language Engineering},
  address   = {New York},
  pages     = {152--166},
  year      = {2019}
}

@misc{JanusGraph,
  title        = {JanusGraph},
  author       = {{JanusGraph Authors}},
  year         = {2024},
  url          = {https://janusgraph.org/},
  note         = {Accessed: 2026-04-28}
}

@misc{AmazonNeptune,
  title        = {Amazon Neptune},
  author       = {{Amazon Web Services, Inc.}},
  year         = {2024},
  url          = {https://aws.amazon.com/neptune/},
  note         = {Accessed: 2026-04-28}
}

@misc{DBEngines,
  title        = {DB-Engines Ranking of Graph DBMS},
  author       = {{solid IT gmbh}},
  year         = {2026},
  url          = {https://db-engines.com/en/ranking/graph+dbms},
  note         = {Accessed: 2026-04-28}
}

@article{Marcos2003,
  author  = {E. Marcos and V. Bel{\'e}n and J. M. Cavero},
  title   = {A Methodological Approach for Object-Relational Database Design using UML},
  journal = {Software and Systems Modeling},
  pages   = {59--72},
  year    = {2003}
}

@inproceedings{De2016,
  author    = {D. Gwendal and S. Gerson and J. Cabot},
  title     = {UMLtoGraphDB: Mapping Conceptual Schemas to Graph Databases},
  booktitle = {Conceptual Modeling},
  publisher = {Springer International Publishing},
  address   = {Cham},
  pages     = {430--444},
  year      = {2016}
}

@article{Pokorny2017,
  author  = {J. Pokorn{\'y} and M. Valenta and J. Kova{\v{c}}i{\v{c}}},
  title   = {Integrity Constraints in Graph Databases},
  journal = {Procedia Computer Science},
  volume  = {109},
  pages   = {975--981},
  year    = {2017}
}

@inproceedings{Wardani2014,
  author    = {D. W. Wardani and J. Kiing},
  title     = {Semantic Mapping Relational to Graph Model},
  booktitle = {2014 International Conference on Computer, Control, Informatics and Its Applications (IC3INA)},
  address   = {Bandung, Indonesia},
  pages     = {160--165},
  year      = {2014}
}

@misc{Mattsson2020,
  author       = {L. Mattsson},
  title        = {Implementing the GraphQL Interface on Top of a Graph Database},
  institution  = {Link{\"o}ping University},
  year         = {2020}
}

@inproceedings{Angles2018,
  author    = {R. Angles},
  title     = {The Property Graph Database Model},
  booktitle = {AMW},
  year      = {2018}
}

@misc{GraphQLQueries,
  title        = {Queries and Mutations},
  author       = {{The GraphQL Foundation}},
  year         = {2024},
  url          = {https://graphql.org/learn/queries/},
  note         = {Accessed: 2026-04-28}
}

@misc{ApolloSchema,
  title        = {Schema Naming Conventions},
  author       = {{Apollo Graph Inc.}},
  year         = {2024},
  url          = {https://www.apollographql.com/docs/technotes/TN0002-schema-naming-conventions/},
  note         = {Accessed: 2026-04-28}
}

@misc{Neo4jGraphQL,
  title        = {Neo4j GraphQL Library},
  author       = {{Neo4j, Inc.}},
  year         = {2024},
  url          = {https://neo4j.com/docs/graphql/current/},
  note         = {Accessed: 2026-04-28}
}

@misc{GremlinProjectStep,
  title        = {Gremlin Project Step},
  author       = {{The Apache Software Foundation}},
  year         = {2024},
  url          = {https://tinkerpop.apache.org/docs/current/reference/#project-step},
  note         = {Accessed: 2026-04-28}
}

@misc{Python,
  title        = {Python},
  author       = {{Python Software Foundation}},
  year         = {2024},
  url          = {https://www.python.org/},
  note         = {Accessed: 2026-04-28}
}

@misc{GremlinPython,
  title        = {gremlinpython (PyPI)},
  author       = {{Python Software Foundation}},
  year         = {2024},
  url          = {https://pypi.org/project/gremlinpython/},
  note         = {Accessed: 2026-04-28}
}

@article{harper2015movielens,
  author    = {F. Maxwell Harper and Joseph A. Konstan},
  title     = {The MovieLens Datasets: History and Context},
  journal   = {ACM Transactions on Interactive Intelligent Systems (TiiS)},
  volume    = {5},
  number    = {4},
  pages     = {19:1--19:19},
  year      = {2015},
  month     = {December},
  doi       = {10.1145/2827872},
  publisher = {ACM}
}

\appendix

\section{Generated GraphQL Schema for the MovieLens Benchmark}
\label{app:schema}

The following listing shows the complete GraphQL schema generated by Graphify for the MovieLens~100K dataset.
It defines four vertex types (\texttt{UserVertex}, \texttt{MovieVertex}, \texttt{GenreVertex}, \texttt{OccupationVertex}), their associated edge types, and the full set of filter, ordering, and pagination input types that the transpiler exposes.

\begin{lstlisting}[
    language=HTML,
    basicstyle=\footnotesize\ttfamily,
    columns=flexible,
    tabsize=2,
    breaklines=true,
    frame=single
]
type Query {
  genre(id: ID!): GenreVertex
  genreList(where: GenreVertexLogicInput,
            orderBy: [GenreVertexOrderByInput!],
            pagination: PaginationInput): [GenreVertex!]!

  occupation(id: ID!): OccupationVertex
  occupationList(where: OccupationVertexLogicInput,
                 orderBy: [OccupationVertexOrderByInput!],
                 pagination: PaginationInput): [OccupationVertex!]!

  user(id: ID!): UserVertex
  userList(where: UserVertexLogicInput,
           orderBy: [UserVertexOrderByInput!],
           pagination: PaginationInput): [UserVertex!]!

  movie(id: ID!): MovieVertex
  movieList(where: MovieVertexLogicInput,
            orderBy: [MovieVertexOrderByInput!],
            pagination: PaginationInput): [MovieVertex!]!
}

type Mutation {
  addGenreVertex(data: GenreVertexInput!): ID!
  updateGenreVertex(id: ID!, data: GenreVertexInput!): ID!
  addOccupationVertex(data: OccupationVertexInput!): ID!
  updateOccupationVertex(id: ID!, data: OccupationVertexInput!): ID!
  addUserVertex(data: UserVertexInput!): ID!
  updateUserVertex(id: ID!, data: UserVertexInput!): ID!
  addMovieVertex(data: MovieVertexInput!): ID!
  updateMovieVertex(id: ID!, data: MovieVertexInput!): ID!
  deleteVertex(id: ID!): ID!
  connectMovieToGenreViaHasGenreEdge(
      source_movie_id: ID!, target_genre_id: ID!): ID!
  connectUserToMovieViaRatedEdge(
      source_user_id: ID!, target_movie_id: ID!,
      data: UserToMovieViaRatedEdgeInput!): ID!
  updateUserToMovieRatedEdge(
      id: ID!, data: UserToMovieViaRatedEdgeInput!): ID!
  connectUserToOccupationViaWorksAsEdge(
      source_user_id: ID!, target_occupation_id: ID!): ID!
  deleteEdge(id: ID!): ID!
}

# --- Vertex types ---

type GenreVertex implements GraphElement {
  id: ID!  label: String!  genreId: Int!  name: String!
  hasGenreIn(whereVertex: MovieVertexLogicInput,
             orderByVertex: [MovieVertexOrderByInput!],
             pagination: PaginationInput): [GenreToMovieHasGenreEdge!]!
}

type OccupationVertex implements GraphElement {
  id: ID!  label: String!  occupationId: Int!  name: String!
  worksAsIn(whereVertex: UserVertexLogicInput,
            orderByVertex: [UserVertexOrderByInput!],
            pagination: PaginationInput): [OccupationToUserWorksAsEdge!]!
}

type UserVertex implements GraphElement {
  id: ID!  label: String!  userId: Int!
  age: Int!  gender: String!  zipCode: Int!
  ratedOut(whereVertex: MovieVertexLogicInput,
           orderByVertex: [MovieVertexOrderByInput!],
           whereEdge: UserToMovieRatedEdgeLogicInput,
           orderByEdge: [UserToMovieRatedEdgeOrderByInput!],
           pagination: PaginationInput): [UserToMovieRatedEdge!]!
  worksAsOut(whereVertex: OccupationVertexLogicInput,
             orderByVertex: [OccupationVertexOrderByInput!],
             pagination: PaginationInput): [UserToOccupationWorksAsEdge!]!
}

type MovieVertex implements GraphElement {
  id: ID!  label: String!  movieId: Int!
  title: String!  releaseDate: String  imdbUrl: String
  hasGenreOut(whereVertex: GenreVertexLogicInput,
              orderByVertex: [GenreVertexOrderByInput!],
              pagination: PaginationInput): [MovieToGenreHasGenreEdge!]!
  ratedIn(whereVertex: UserVertexLogicInput,
          orderByVertex: [UserVertexOrderByInput!],
          whereEdge: UserToMovieRatedEdgeLogicInput,
          orderByEdge: [UserToMovieRatedEdgeOrderByInput!],
          pagination: PaginationInput): [MovieToUserRatedEdge!]!
}

# --- Edge types ---

type MovieToGenreHasGenreEdge implements GraphElement {
  id: ID!  label: String!  genre: GenreVertex! }
type GenreToMovieHasGenreEdge implements GraphElement {
  id: ID!  label: String!  movie: MovieVertex! }
type UserToMovieRatedEdge implements GraphElement {
  id: ID!  label: String!  rating: Int!
  timestamp: String!  movie: MovieVertex! }
type MovieToUserRatedEdge implements GraphElement {
  id: ID!  label: String!  rating: Int!
  timestamp: String!  user: UserVertex! }
type UserToOccupationWorksAsEdge implements GraphElement {
  id: ID!  label: String!  occupation: OccupationVertex! }
type OccupationToUserWorksAsEdge implements GraphElement {
  id: ID!  label: String!  user: UserVertex! }

# --- Input types (mutations) ---

input GenreVertexInput      { genreId: Int!  name: String! }
input OccupationVertexInput { occupationId: Int!  name: String! }
input UserVertexInput  { userId: Int!  age: Int!  gender: String!  zipCode: Int! }
input MovieVertexInput { movieId: Int!  title: String!
                         releaseDate: String  imdbUrl: String }
input UserToMovieViaRatedEdgeInput { rating: Int!  timestamp: String! }

# --- Filter input types (excerpt: UserVertexLogicInput shown as representative) ---

input UserVertexLogicInput {
  userId_EQ: Int    userId_NEQ: Int    userId_GT: Int    userId_GTE: Int
  userId_LT: Int    userId_LTE: Int
  age_EQ: Int       age_NEQ: Int       age_GT: Int       age_GTE: Int
  age_LT: Int       age_LTE: Int
  gender_EQ: String  gender_NEQ: String  gender_GT: String  gender_GTE: String
  gender_LT: String  gender_LTE: String
  zipCode_EQ: Int   zipCode_NEQ: Int   zipCode_GT: Int   zipCode_GTE: Int
  zipCode_LT: Int   zipCode_LTE: Int
  OR: [UserVertexLogicInput!]   AND: [UserVertexLogicInput!]
}
# Analogous filter inputs exist for GenreVertex, OccupationVertex,
# MovieVertex, and the UserToMovieRatedEdge type.

# --- Ordering and pagination ---

enum OrderDirection { ASC  DESC }
input PaginationInput { offset: Int!  limit: Int! }

input UserVertexOrderByInput {
  property: UserVertexProperty!  order: OrderDirection! }
enum UserVertexProperty { userId  age  gender  zipCode }
# Analogous orderBy inputs and enums exist for all other vertex/edge types.

interface GraphElement { id: ID!  label: String! }
\end{lstlisting}

\section{Benchmark Queries}
\label{app:queries}

The four queries used in the empirical evaluation (Section~\ref{sec:complexity}) are listed below in the order they appear in Table~\ref{tab:queries}.

\begin{lstlisting}[
    language=HTML,
    basicstyle=\footnotesize\ttfamily,
    columns=flexible,
    tabsize=2,
    breaklines=true,
    frame=single,
    caption={SimpleLookup --- movie lookup by exact title ($S=3$, $W=1$)},
    label={lst:simple}
]
query SimpleLookup {
  movieList(where: { title_EQ: "Toy Story (1995)" }) {
    id
    releaseDate
    imdbUrl
  }
}
\end{lstlisting}

\begin{lstlisting}[
    language=HTML,
    basicstyle=\footnotesize\ttfamily,
    columns=flexible,
    tabsize=2,
    breaklines=true,
    frame=single,
    caption={ComplexFilter --- user list with compound filter, ordering, and pagination ($S=2$, $W=2$, $K=1$)},
    label={lst:complex}
]
query ComplexFilter {
  userList(
    where: { AND: [{ age_GT: 18 }, { gender_EQ: "M" }] }
    orderBy: [{ property: age, order: DESC }]
    pagination: { offset: 0, limit: 10 }
  ) {
    userId
    age
  }
}
\end{lstlisting}

\begin{lstlisting}[
    language=HTML,
    basicstyle=\footnotesize\ttfamily,
    columns=flexible,
    tabsize=2,
    breaklines=true,
    frame=single,
    caption={UserRatings --- single user with one-hop traversal to rated movies ($S=5$, $D=1$)},
    label={lst:userratings}
]
query UserRatings {
  user(id: "<user_graph_id>") {
    userId
    ratedOut(pagination: { offset: 0, limit: 5 }) {
      rating
      movie {
        title
      }
    }
  }
}
\end{lstlisting}

\begin{lstlisting}[
    language=HTML,
    basicstyle=\footnotesize\ttfamily,
    columns=flexible,
    tabsize=2,
    breaklines=true,
    frame=single,
    caption={GenreDemographics --- genre with three-hop nested traversal ($S=10$, $W=1$, $D=3$)},
    label={lst:genre}
]
query GenreDemographics {
  genre(id: "<genre_graph_id>") {
    name
    hasGenreIn(pagination: { offset: 0, limit: 3 }) {
      movie {
        title
        ratedIn(
          whereEdge: { rating_GTE: 4 }
          pagination: { offset: 0, limit: 5 }
        ) {
          user {
            age
            worksAsOut {
              occupation {
                name
              }
            }
          }
        }
      }
    }
  }
}
\end{lstlisting}

\end{document}